%% file: compact-loop-paper-v3.tex
\newcommand{\version}{February 28, 2013}
\documentclass[11pt,a4paper,english,pdftex]{article}

\usepackage[bindingoffset=0.3cm,textheight=22.5cm,hdivide={2.5cm,*,2.5cm}, vdivide={*,22cm,*}]{geometry}
\usepackage[pdftex,bookmarksnumbered=true,breaklinks=true]{hyperref}

\usepackage{amsmath,amsfonts,amssymb,slashed,mathrsfs}
\usepackage[pdftex]{graphicx}

\usepackage[numbers,square,comma,sort&compress]{natbib}

\IfFileExists{dsfont.sty}
	{\usepackage{dsfont}
         \newcommand{\id}{\mathds{1}}}
	{\typeout{Package dsfont.sty was not found, using alternative macros.}
         \let\mathds=\mathbb
         \newcommand{\id}{\mbox{1 \kern-.59em {\rm l}}}}
\let\one=\id

\makeatletter

         \@addtoreset{equation}{section}
\makeatother

\newcommand{\nocontentsline}[3]{}
\newcommand{\tocless}[3]{\bgroup\let\addcontentsline=\nocontentsline#1{#2}#3\egroup}

\newcommand{\qed}{\nobreak \ifvmode \relax \else
      \ifdim\lastskip<1.5em \hskip-\lastskip
      \hskip1.5em plus0em minus0.5em \fi \nobreak
      \vrule height0.75em width0.5em depth0.25em\fi}

\newcommand{\be}{\begin{equation}}
\newcommand{\ee}{\end{equation}}
\newcommand{\eq}[1]{(\ref{#1})}

\def\nn{\nonumber}
\def\bea{\begin{eqnarray}}
\def\eea{\end{eqnarray}}
\def\obar{\overline}

\def\beqa{\begin{eqnarray}} 
\def\eeqa{\end{eqnarray}} 
\def\beq{\begin{equation}} 
\def\eeq{\end{equation}}

\def\Tr{{\rm Tr}}

\def\a{\alpha}          
\def\b{\beta}           
\def\c{\gamma}  

\def\l{\lambda} \def\L{\Lambda} \def\la{\lambda}
\def\X{\Xeta}

 \def\cH{{\cal H}} 
 \def\cK{{\cal K}} 
\def\cM{{\cal M}} \def\cN{{\cal N}} \def\cO{{\cal O}}
  
  \def\cU{{\cal U}}

\newcommand{\R}{\mathds{R}}

\newcommand{\Z}{\mathds{Z}}

\def\bit{\begin{itemize}}
\def\eit{\end{itemize}}

\def\({\left(}
\def\){\right)}
\def\diag{\mbox{diag}}

\def\bcomment#1{}

\renewcommand{\Box}{\square}

\newcommand{\nc}{non-com\-mu\-ta\-tive}

\newcommand{\figref}[1]{Fig.~\ref{#1}}			
\newcommand{\tinv}[1]{\tfrac{1}{#1}}
\newcommand{\co}[2]{[#1,#2]}						
\renewcommand{\a}{\alpha}
\renewcommand{\b}{\beta}
\newcommand{\g}{\gamma}
\renewcommand{\d}{\delta}
\newcommand{\e}{\epsilon}

\renewcommand{\th}{\theta}

\renewcommand{\l}{\lambda}
\newcommand{\m}{\mu}

\newcommand{\s}{\sigma}

\newcommand{\G}{\Gamma}
\newcommand{\vG}{\varGamma}

\newcommand{\Th}{\Theta}

\renewcommand{\L}{\Lambda}

\renewcommand{\Xi}{\Xi}

\title{\begin{flushright}
       \small{UWThPh-2013-05}\\{LA-UR-13-21320}
       \end{flushright}
\vspace{3em}
\bf
Compactified rotating branes in the matrix model, and excitation spectrum towards one loop}
\author{Daniel N. Blaschke\footnote{dblaschke@lanl.gov}~, Harold C. Steinacker\footnote{harold.steinacker@univie.ac.at}}

\date{\version}
\begin{document}

\maketitle

\begin{center}
\renewcommand{\thefootnote}{\fnsymbol{footnote}}
\vspace{-0.3cm}\footnotemark[1]\textit{Los Alamos National Laboratory, Theory Division\\Los Alamos, NM, 87545, USA}\\[0.3cm]
\footnotemark[2]\textit{University of Vienna, Faculty of Physics\\
Boltzmanngasse 5, A-1090 Vienna, Austria}
\vspace{0.5cm}
\end{center}%
\begin{abstract}

We study compactified brane solutions of type $\R^4 \times K$ in the IIB matrix model, and
obtain explicitly the bosonic and fermionic fluctuation spectrum required to compute the one-loop 
effective action. 
We verify that the one-loop contributions are UV finite for $\R^4 \times T^2$, and supersymmetric
for $\R^3 \times S^1$. The higher Kaluza-Klein modes are shown to have a gap 
in the presence of flux on $T^2$, and potential problems concerning stability are discussed.

\end{abstract}

\newpage
\tableofcontents

\section{Introduction --- the IKKT or IIB model}
\label{sec:IKKT-IIB}

In recent years matrix models, notably the IKKT model~\cite{Ishibashi:1996xs}, have become increasingly promising candidates to 
describe quantum theories of matter and gravity.
The model describes dynamical branes, which support (non-commutative) gauge theory on their
world-volume. By considering solutions with compactified extra dimensions, the low-energy effective theory
on these branes
can acquire non-trivial structure, and may become physically relevant \cite{Chatzistavrakidis:2011gs}, 
cf. \cite{Aoki:2010gv}.
Such solutions  with geometry 
$\R^4 \times K \subset \R^{10}$ were found recently in \cite{Steinacker:2011wb}, generalizing similar solutions for the BFSS 
model \cite{Bak:2001kq}. In contrast to the well-known 10-dimensional compactifications \cite{Connes:1997cr}, they have the advantage that 
the one-loop effective action (and presumably also for higher loops) is finite. 
Since this is a very important point in favour of the present type of solutions, 
we elaborate in this short note the basic ingredients for the one-loop computation.
Furthermore, gravity is expected to be an emergent force on such branes --- for a general introduction to the subject, 
the interested reader is referred to~\cite{Steinacker:2010rh}, 
and to~\cite{Steinacker:2012ra,Steinacker:2012ct} for more specific aspects in the present context.

In this paper, we follow up on our recent results on the one-loop effective action~\cite{Blaschke:2010rr,Blaschke:2011qu}
on such branes,
adapted to the present context.
In particular, we consider compactified brane solutions of type $\R^3 \times S^1$
 and $\R^4 \times T^2$ in the IKKT  model, and elaborate some 
relevant aspects of the one-loop quantum corrections for the present solutions. 
The point is that although UV finiteness is a general consequence of the maximal supersymmetry of the 
matrix model, the specific one-loop effective action does depend on the background, which may or may not preserve
some supersymmetry. It turns out that the simplest $\R^2 \times S^1$ brane is indeed supersymmetric, and the 
one-loop correction to the effective action vanishes identically. In contrast, the  $\R^4 \times T^2$ solutions
turn out not to preserve any supersymmetry, and the one-loop effective action is finite but non-trivial.

We start with a general discussion of the IKKT (or type IIB) matrix model and its properties 
in the context of emergent gravity, and  explain how to compute the effective one-loop action. 
The IKKT or IIB model \cite{Ishibashi:1996xs} is defined by the following action
\begin{align}
S_{\rm IKKT}&=-(2\pi)^2\Tr\left(\co{X^a}{X^b}\co{X_a}{X_b}\,\, + \, 2\obar\Psi \gamma_a[X^a,\Psi] \right) 
\,,
\label{IKKT-MM}
\end{align}
where $X^a,\,\, a= 0,1,2,\ldots,9$ are Hermitian matrices,
$\Psi$ is a matrix-valued Majorana-Weyl spinor of $SO(9,1)$,
and the $\g_a$ form the corresponding Clifford algebra.
The model is obtained by dimensional reduction of the 10-dimensional $SU(N)$ Super-Yang-Mills theory
to a point, and taking the $N \to\infty$ limit. 
Indices are raised and lowered using the fixed background metric  $g_{ab} = \eta_{ab}$.
This action is invariant under the following  symmetries:
\be
\begin{array}{lllll}
X^a \to U^{-1} X^a U\,,\quad  &\Psi \to U^{-1} \Psi U\,,\quad  &  U \in U(\cH)\,\quad  & \mbox{gauge invariance,}  \\
X^a \to \L(g)^a_b X^b\,,\quad  &\Psi_\a \to \tilde \pi(g)_\a^\b \Psi_\b\,,\quad  & g \in \widetilde{SO}(9,1)\,\; 
  & \mbox{Lorentz symmetry,}  \\
X^a \to X^a + c^a \one\,,\quad & \quad & c^a \in \R\,\quad  & \mbox{translational symmetry,}
\end{array}
\label{transl-inv}
\ee
where the tilde indicates the corresponding spin group, as well as 
the $\cN=2$ matrix supersymmetry \cite{Ishibashi:1996xs}
\begin{align}
\d_{\e}^{(1)}\psi&=\frac i2[X_a,X_b]\G^{ab}\e \,, 
&\d_{\e}^{(1)}X_a&=i\bar\e\G_a\psi\,, \nn\\
\d_{\xi}^{(2)}\psi&=\xi\,,
&\d_{\xi}^{(2)}X_a&=0
\,. \label{susy}
\end{align}
Here matrices are identified with operators on a separable Hilbert space $\cH$, and $U(\cH)$
denotes the group of unitary operators resp. matrices on $\cH$.

Although an explicit proof to all orders is still missing, there are good reasons to believe that this model 
is in fact UV finite on a 4-dimensional\footnote{This is expected to hold also in the presence of 
compactified extra dimensions, due to the non-commutativity of the brane.} background~\cite{Jack:2001cr}. 
Furthermore, numerical evidence for the emergence of 3+1-dimensional space-time within the IKKT model has been obtained 
recently in Ref.~\cite{Kim:2011cr}, providing further motivation to study the effective physics of 4-dimensional backgrounds. 
Additionally, it was shown in~\cite{Chatzistavrakidis:2011gs} how the particle spectrum of the standard model may be 
correctly reproduced from that action by considering specific brane solutions, hence showing how realistic physics 
may emerge from the IKKT model.

\subsection*{One-loop action}

In order to introduce a common notation which works for all fields, let 
\begin{align}
\begin{array}{rlcl}(\Sigma_{ab}^{(\psi)})^\a_\b &= \frac i4 [\gamma_a,\gamma_b]^\a_\b \,, && \mbox{fermions,}  \\
                      (\Sigma_{ab}^{(Y)})^c_d &= i(\d^c_a g_{bd} - \d^c_b g_{ad}) \,, && \mbox{bosonic matrices,} \\
                 \Sigma_{ab}^{(c)} &= 0 \,, && \mbox{ghosts,}
                    \end{array}  
\end{align}
be the generators of $SO(D)$ on the spinor and vector irreducible representations 
($[\gamma_a,\gamma_b]$ is understood to be the irreducible chiral representation).  
The real part of the 1-loop contribution for the  IKKT model can then be written as 
\cite{Blaschke:2011qu} (cf. \cite{Ishibashi:1996xs})
\begin{align}
\vG:=\Gamma^{\rm real}_{{\rm 1-loop}}[X] &= - \frac 12 \Tr \int\limits_0^\infty \frac {d\a}{\a} 
   \Big( e^{-\a(\Box  + \Sigma^{(Y)}_{ab}[\Th^{ab},.])}
  - \frac 12 e^{-\a(\Box  + \Sigma^{(\psi)}_{ab}[\Th^{ab},.]) }
 - 2 e^{-\a\Box}  \Big) 
\,, \label{Gamma-schwinger-susy}
\end{align}
which is manifestly $SO(10)$ invariant. 
Here 
\begin{align}
 \Box = [X^a,[X_a,.]]
\end{align}
is the matrix Laplace operator.
As pointed out before, there may be an additional imaginary contribution to the action
for some backgrounds, which can be 
interpreted as Wess-Zumino term related to a global anomaly 
of the chiral $SU(4)$ R-symmetry. An analogous formula applies for the $D$-dimensional reduced model.

As a first check, we note that $\vG[X]$ vanishes 
in the free case (i.e. for an unperturbed $\R^4_\theta$ background with $\Theta^{ab} \sim \one_\cH$), 
provided the  $D-2$ effective degrees of freedom from (bosons $-$ ghosts)
cancel with the fermions. This holds true in the $D=10$ IKKT model for Majorana-Weyl fermions.

The above formula \eq{Gamma-schwinger-susy} is valid for arbitrary backgrounds.
It suggests the following expansion:
\begin{align}
\vG[X]\! &:= - \frac 12 \Tr \int_0^\infty \frac {d\a}{\a} 
   \Big( e^{-\a(\Box  + \Sigma^{(Y)}_{ab}[\Th^{ab},.])}
  - \frac 12 e^{-\a(\Box  + \Sigma^{(\psi)}_{ab}[\Th^{ab},.]) }
 - 2 e^{-\a\Box}  \Big) \nn\\
 &= \frac 12 \Tr \Bigg(\sum_{n>0} \frac{(-1)^{n+1}}n \Big((\Sigma^{(Y)}_{ab}\Box^{-1}[\Th^{ab},.])^n 
  \, -\frac 12 (\Sigma^{(\psi)}_{ab}\Box^{-1}[\Th^{ab},.])^n \Big)  \Bigg) \nn\\
 &= \frac 12 \Tr \Bigg(\!\! -\frac 14 (\Sigma^{(Y)}_{ab} \Box^{-1}[\Th^{ab},.] )^4 
 +\frac 18 (\Sigma^{(\psi)}_{ab} \Box^{-1}[\Th^{ab},.])^4 \,\, +  \cO(\Box^{-1}[\Th^{ab},.])^5 \! \Bigg) .
\label{Gamma-IKKT}
\end{align}
Moreover, the $SO(10)$ resp. $SO(9,1)$ symmetry is still manifest. 
These statements are independent of any specific background, and reflect the maximal SUSY of the model.
Therefore the traces  behave
as $\int\! d^{2n} p\, \frac 1{p^8}$ in the UV on $2n$-dimensional backgrounds, which is convergent
for $2n < 8$.
In particular, the model is one-loop finite\footnote{One-loop finiteness
holds even on 6D backgrounds, but not for higher loops. This is consistent with well-known
results \cite{Green:1982sw,Fradkin:1983jc,Howe:1983jm} for SYM theory.}  on 4-dimensional backgrounds.

\section{Compactified brane solutions and split non-commutativity}

There is an interesting class of solutions $\cM = \cM^4\times \cK$ with compactified (``spinning'') extra dimensions 
stabilized by angular momentum. These solutions exist only in the case of Minkowski signature. 
At the semi-classical level, the basic feature is that the Poisson structure relates the compact 
with the non-compact space, i.e. 
\be
\{x^\mu,y^i\} \neq 0
\,, \label{split-NC}
\ee 
where $x^\mu$ are coordinates on $\cM^4$ and $y^i$ are coordinates on $\cK$. 
This will be indicated by the name ``split non-commutativity''.
If  $\cK$ is 4-dimensional, then one might even have commutative $\cM^4$, so that $\{x^\mu, x^\nu\} =0$. 
A standard example is the canonical symplectic structure on the cotangent bundle $T^* \cK$. 
Such a structure is indispensable for solutions $\cM^4\times \cK$ with compact  $\cK$, since
there is no harmonic embedding  of compact spaces. The effective signature on $\cM^4$ 
can have Minkowski signature. 
Several examples of such solutions of the matrix model have been given, including 
$\cK= T^2,\,   \cK = S^1 \times S^3$ and $\cK = S^2 \times S^2$ \cite{Steinacker:2011wb}.
Here we focus  on solutions with $\cK = S^1$ and $\cK = T^2$.

\subsection{The fuzzy cylinder}
\label{sec:fuzzy-cylinder}

The fuzzy cylinder \cite{Janssen:2004jz,Chaichian:1998kp,Bak:2001kq}
$S^1 \times_\xi \R$ is defined in terms of 3 Hermitian matrices which satisfy
\begin{align}
[X^1,X^3] &= i \xi X^2, & [X^2,X^3] &= - i \xi X^1,  \nn\\
 (X^1)^2 + (X^2)^2 &= R^2, & [X^1,X^2] &= 0  .
\end{align}
Defining $U := X^1+iX^2$ and  $U^\dagger := X^1-iX^2$,
this can be stated more transparently as 
\begin{align}
U U ^\dagger &= U^\dagger U = R^2     \nn\\
\, [U, X^3] &= \xi U\,,  & [U^\dagger, X^3] &= - \xi U^\dagger  \,.
 \label{fuzzy-cylinder} 
\end{align}
This algebra has the following 
irreducible representation
\begin{align}
U |n\rangle &= R   |n + 1\rangle , \qquad   U^\dagger  |n\rangle = R  |n - 1\rangle  \nn\\
X^3 |n\rangle &= \xi n  |n\rangle,  \qquad \qquad n \in \Z, \,\, \xi \in \R
\label{cylinder-rep}
\end{align}
on a Hilbert space $\cH$, where $|n\rangle$ form an orthonormal basis.

We will always assume that the non-compact directions $\R^4$ are embedded along the $0,1,2,3$ directions.
We focus on the two following solutions:

\subsection{Rotating cylinder}

Now consider a solution of type $\R^3\times S^1 \subset \R^5$ with the embedding 
\begin{align}
X^A &= \begin{pmatrix}
X^0\\X^1\\X^2\\R U
\end{pmatrix}
\,, 
&\textrm{with} \qquad 
\co{X^\mu}{X^\nu}&=i\th^{\mu\nu}\,, &
\co{X^\mu}{U}&=-\xi^\mu U\,  
,
\end{align}
where the embedding metric of $\R^3$ has Minkowski signature.
Such operators can be easily realized by combining a fuzzy cylinder  with a 
Moyal-Weyl quantum plane $\R^2_\theta$, multiplying the unitary operator with a 
suitable phase factor $e^{i \tilde p_\mu X^\mu}$ \cite{Steinacker:2011wb}.
Functions are represented by\footnote{Note that $\cM = \R^3 \times S^1$ is symplectic while $\R^3$ is 
not. Thus one can raise or lower indices by introducing an
effective metric $G^{\mu\nu}$ on $\cM$ following \cite{Steinacker:2010rh}. 
In this paper we short-cut these
geometrical considerations, and use a tilde to indicate that a momentum is in 
$T^*_\cM$ rather than $T_\cM$.}
\begin{align}
f(U,X)&=\sum_n\int\! d^3\tilde p f_n(\tilde p) e^{i\tilde p_\m X^\m}U^n
\,. 
\end{align}
We will need the following commutators:
\begin{align}
\co{X^\mu}{e^{i\tilde p X}U^n}&= -(\theta^{\mu\nu} \tilde p_\nu + n\xi^\mu) e^{i\tilde p X}U^n\,,\nn\\
\co{U}{e^{i\tilde p X}U^n}&=\left(e^{i\tilde p_\mu\xi^\mu}-1\right)e^{i\tilde p X}U^{n+1}\,,\nn\\
\co{U^\dagger}{e^{i\tilde p X}U^n}&=\left(e^{-i\tilde p_\mu\xi^\mu}-1\right)e^{i\tilde p X}U^{n-1}
\,.
\end{align}
We introduce a basis of wave functions as
\begin{align}
|n,p\rangle &= e^{i \tilde p_\mu (X^\mu - \xi^\mu/2)}  U^{-n} \,, \nn\\
\co{U}{.} |n,p\rangle &= (e^{i \tilde p_\mu\xi^\mu}-1) U |n,p\rangle  = 2\sin\Big(\frac{\tilde p_\mu\xi^\mu}2\Big) |n-1,p\rangle \,, \nn\\
\co{U^\dagger}{.} |n,p\rangle &= -(e^{-i \tilde p_\mu\xi^\mu}-1) U^\dagger |n,p\rangle  = 2\sin\Big(\frac{\tilde p_\mu\xi^\mu}2\Big) |n+1,p\rangle
\label{basis-np}
\, . 
\end{align}
Then the Laplacian of a general function becomes
\begin{align}
\Box |n,p\rangle &= \Big(\co{R {U}^\dagger}{\co{R {U}}{ . }}+\co{X^\m}{\co{X_\m}{ . }}\Big)|n,p\rangle \nn\\
 &= \l_{n,p} |n,p\rangle,  \nn\\
\l_{n,p}&=4R^2\sin^2\left(\frac{\tilde p_\mu\xi^\mu}2\right)+ (p-n\xi)\cdot (p-n\xi) \,, \nn\\
(p-n\xi)\cdot (p-n\xi) &= (p^\mu - n\xi^\mu) (p^\eta - n\xi^\eta) \eta_{\mu\eta}\,   \nn\\
p^\mu &= \theta^{\mu\nu} \tilde p_\nu
\,. \label{eq:thematrix_newbox}
\end{align}
Now the on-shell condition $\Box {U}=0$ is solved if $\xi\cdot \xi = 0$, 
and $\Box X^\mu = 0$ always,
i.e. $\xi^\mu$ is light-like w.r.t. $\eta_{\mu\nu}$.
This leads to an important difference to the basic fuzzy cylinder: the quadratic terms in $n$ cancel, and
\begin{align}
 \l_{n,p}
 &= 4R^2\sin^2\left(\frac{\tilde p_\mu\xi^\mu}2\right)+ p \cdot p - 2 n p \cdot \xi   
 \,. 
\end{align}
For the higher Kaluza-Klein modes, this amounts to a  dispersion relation where the origin in 
momentum space is shifted.

\subsubsection*{Eigenvalues and one-loop computation.}

To fix the conventions, let us embed $S^1\subset \R^2$ with coordinates $y^2,y^3$, hence
\begin{align}
 U = Y^2+iY^3
\end{align}
and $\R^2$ with Minkowski signature and coordinates $x^0,x^1$. 
For simplicity we drop the $x^2$ coordinate, assuming that $\xi^2=0$.
Furthermore, $\xi^0 = -\xi^1$ ensures $\xi\cdot\xi = 0$.
Hence
\begin{align}
 [U,(x^0-x^1)] &= \xi U\,,   & 
 [U,(x^0+x^1)] &= 0
\,, \label{convention-lightlike}
\end{align}
where $\xi = \xi^0 - \xi^1$.

\paragraph{Fermionic modes.}

Let us choose the 4-dimensional chiral representation of the Gamma matrices:
\begin{align}
\Sigma^{\psi}_{ab} &=\frac{i}{4}\co{\g_a}{\g_b} \,
                   = \begin{pmatrix}
                      \Sigma_{ab}^{(+)} & 0\\
                      0 & \Sigma_{ab}^{(-)}
                     \end{pmatrix}  , \nn\\
\Sigma^{(\psi)}_{0i}&= \frac 1{2i}\begin{pmatrix}
                                  \s_i & 0 \\
                                  0 & -\s_i
                                  \end{pmatrix},   &
\Sigma^{(\psi)}_{ij}&= \frac 1{2}\e_{ijk}\begin{pmatrix}
                                  \s_k & 0 \\
                                  0 & \s_k
                                  \end{pmatrix}
. 
\end{align}
We will write $\g_\pm=\tinv{2}\left(\g_1\mp i\g_2\right)$,
hence $\Sigma^{(\psi)}_{0,\pm}=\frac{i}{8}\co{\g_0}{\g_1\mp i\g_2}$.
For both $\Sigma^{(Y)}_{ab}$ and $\Sigma^{(\psi)}_{ab}$ we have
\begin{align}
\Sigma_{02}\Th^{02}+\Sigma_{03}\Th^{03}=\Sigma_{0+}\Th^{0+}+\Sigma_{0-}\Th^{0-}
\,, 
\end{align}
hence
\begin{align}
\Sigma_{0\pm}^{(\psi)}&=\tinv{2}\left(\Sigma_{02}\mp i\Sigma_{03}\right)
 = \frac 1{4i}\begin{pmatrix}
                                  (\s_2\mp i\s_3) & 0 \\
                                  0 & -(\s_2\mp i\s_3)
                                  \end{pmatrix}
\,, 
\end{align} 
and similarly 
\begin{align} 
\Sigma_{1\pm}^{(\psi)}&=\tinv{2}\left(\Sigma_{12}\mp i\Sigma_{13}\right) 
 = \frac 1{4}\begin{pmatrix} 
                                  (\s_3 \pm i \s_2) & 0 \\ 
                                  0 & (\s_3 \pm i \s_2)
                                  \end{pmatrix}
\, = \pm\frac i{4}\begin{pmatrix}
                                  (\s_2\mp i\s_3) & 0 \\
                                  0 & (\s_2\mp i\s_3)
                                  \end{pmatrix}
\,. 
\end{align}
Now recalling that $U=Y^2+iY^3$, we have
\begin{align}
\Sigma^{\psi}_{ab}\Theta^{ab} &=  \Sigma_{0\pm}^{(\psi)} \Theta^{0\pm} + \Sigma_{1\pm}^{(\psi)} \Theta^{1\pm} \nn\\
 &= \frac 12 (\Sigma_{0\pm}^{(\psi)} - \Sigma_{1\pm}^{(\psi)})( \Theta^{0\pm} - \Theta^{1\pm}) 
 + \frac 12 (\Sigma_{0\pm}^{(\psi)} + \Sigma_{1\pm}^{(\psi)})( \Theta^{0\pm} + \Theta^{1\pm})  \nn\\
 &= -\frac \xi2 (\Sigma_{0+}^{(\psi)} - \Sigma_{1+}^{(\psi)}) R U
  + \frac \xi2 (\Sigma_{0-}^{(\psi)} - \Sigma_{1-}^{(\psi)}) R U^\dagger \nn\\
 &= -\frac{R\xi}2 \frac 1{2i}\begin{pmatrix}
                                  (\s_2- i\s_3)U   & 0 \\
                                  0 & - (\s_2+ i\s_3) U^\dagger
                           \end{pmatrix} 
\,, 
\end{align}
since 
\begin{align}
i\Theta^{+,+} &= [X^0+X^1,RU] = 0\,, \qquad i\Theta^{-,+} = [X^0-X^1,RU] = - R\xi U \,, \nn\\
i\Theta^{+,-} &= [X^0+X^1,RU^\dagger] = 0\,, \qquad i\Theta^{-,-} = [X^0-X^1,RU^\dagger] = R\xi U^\dagger 
\,. 
\end{align}
To make things more transparent we use a non-standard representation of the $\sigma$ matrices such that
\begin{align}
 \s_2- i\s_3 &= 2\begin{pmatrix}
                0 & 0\\ 1 & 0
               \end{pmatrix}, \qquad 
\s_2 +i\s_3 = 2\begin{pmatrix}
                0 & 1\\ 0 & 0
               \end{pmatrix},
\nn\\
\Sigma^{\psi}_{ab}\Theta^{ab} 
&= i\frac{R\xi}{2}\begin{pmatrix}
                                \begin{pmatrix}
                                        0 & 0\\ U & 0
                                \end{pmatrix}  & 0 \\
          0 & - \begin{pmatrix}
                0 & U^\dagger \\ 0 & 0
                \end{pmatrix}       
                           \end{pmatrix} .
\end{align}
We note that this matrix  has a non-trivial kernel (with rank 2, one for each chirality), 
i.e. there are 2 spinors $\e$ which satisfy
\begin{align}
 \Sigma^{\psi}_{ab}\Theta^{ab} \e = 0 
 \label{susy-cond}
 \, .
\end{align}
This means that the background is supersymmetric \cite{Ishibashi:1996xs}.

We can now evaluate the fermionic one-loop contribution. Using
the following ansatz for the spinor  
\begin{align}
\psi_{n,p} = \begin{pmatrix}
  a_{n+1} |n+1,p\rangle \\ 
  b_{n} |n,p\rangle \\
  c_{n}  |n,p\rangle  \\
  d_{n-1} |n-1,p\rangle 
 \end{pmatrix},
\end{align}
we obtain using \eq{basis-np}  
\begin{align}
 \l_n M_\psi  &= (\Box + \Sigma^{\psi}_{ab}[\Theta^{ab},.])\psi =  M_\psi \psi \nn\\
 &=  \begin{pmatrix}
     \l_{n+1,p}  & 0 & 0 & 0 \\
     iR\xi\sin(\frac{\tilde p\xi}2) & \l_{n,p} & 0 & 0 \\
     0 & 0 & \l_{n,p} & -iR\xi\sin(\frac{\tilde p\xi}2)  \\
    0 & 0 & 0 & \l_{n-1,p}
    \end{pmatrix} \psi  \, .
  \end{align}
Therefore
\begin{align}
0 &=  \begin{pmatrix}
      \l_{n+1,p} -\l_n & 0 & 0 & 0 \\
     \frac i2 \xi\a_p &\l_{n,p}  -\l_n & 0 & 0 \\
     0 & 0 & \l_{n,p}  -\l_n & -\frac i2 \xi\a_p  \\
    0 & 0 & 0 & \l_{n-1,p}  -\l_n
    \end{pmatrix} \psi  
\,, 
\end{align}
where
\be
\a_p = 2R \sin(\frac{\tilde p_\mu\xi^\mu}2), \qquad 
\l_{n,p} = \a_p^2 +p\cdot p-2n\xi\cdot p 
\,. 
\ee
The vanishing determinant condition yields
\begin{align}
(\l_{n,p}-\la)^2(\l_{n+1,p}-\la)(\l_{n-1,p}-\la) = 0 \,, 
\label{det-vanish-fermions}
\end{align}
so that the eigenvalues are given  by
\begin{align}
\l_1&=\l_2=\a_p^2+p\cdot p-2n\xi\cdot p \,, \nn\\
\l_{3,4}&=\a_p^2+p\cdot p - 2(n\pm1)\left(\xi\cdot p\right)
\,.
\end{align}
This coincides with the eigenvalues of the bosonic sector, as we will show next.

\paragraph{Bosonic representation.}

To evaluate the bosonic one-loop contribution, we note that
\begin{align}
&\Box  + \Sigma^{(Y)}_{ab}[\Th^{ab},.]= \cU^{-1} M_Y \cU \,, \qquad
\cU= \begin{pmatrix}
 1 & 0 & 0 & 0 \\
 0 & 1 & 0 & 0\\
 0 & 0 & 1/\sqrt{2} & i/\sqrt{2}\\
 0 & 0 & -i/\sqrt{2} & -1/\sqrt{2}
\end{pmatrix} \,, 
\end{align}
where
\begin{align}
M_Y = \begin{pmatrix}
 \Box & 0 & -\sqrt{2}\xi_0 R\,\co{U^\dagger}{.} & \sqrt{2}i\xi_0 R\,\co{U}{.} \\
 0 & \Box & -\sqrt{2}\xi_1 R\,\co{U^\dagger}{.} & \sqrt{2}i\xi_1 R\,\co{ U}{.} \\
 \sqrt{2}\xi_0 R\,\co{U}{.} & -\sqrt{2}\xi_1 R\,\co{U}{.} & \Box & 0\\
 \sqrt{2} i\xi_0 R\,\co{U^\dagger}{.}  & -\sqrt{2} i\xi_1 R\,\co{U^\dagger}{.} & 0 & \Box 
\end{pmatrix} .
\end{align}
The eigenvalues of $M_Y$ are obtained from
\begin{align}
\begin{pmatrix}
\l_{n,p}-\l & 0 & -\sqrt{2}\xi_0\a_p & \sqrt{2}i\xi_0\a_p \\
0 & \l_{n,p}-\l & -\sqrt{2}\xi_1\a_p & \sqrt{2}i\xi_1\a_p \\
\sqrt{2}\xi_0\a_p & -\sqrt{2}\xi_1\a_p & \l_{n-1,p}-\l & 0 \\
\sqrt{2}i\xi_0\a_p & -\sqrt{2}i\xi_1\a_p & 0 & \l_{n+1,p}-\l
\end{pmatrix}
\begin{pmatrix}
a_n\\a_n\\b_{n-1}\\c_{n+1}
\end{pmatrix}
=0 \, .
\end{align}
It follows with $\xi\cdot\xi=0=\xi_0^2-\xi_1^2$ that
\begin{align}
(\l_{n,p}-\la)^2(\l_{n+1,p}-\la)(\l_{n-1,p}-\la) = 0 
\,, \label{det-vanish-bosons}
\end{align}
where
\begin{align}
\l_{n,p}=\a_p^2+(p-n\xi)\cdot(p-n\xi) =\a_p^2+p\cdot p-2n\xi\cdot p
\,. 
\end{align}
This coincides with the eigenvalues in the fermionic sector \eq{det-vanish-fermions}, consistent 
with supersymmetry \eq{susy-cond}.
Since the eigenvalues for the bosonic and fermionic modes coincide 
(and also with the ghosts where $\l = \l_{n,p}$), 
the one-loop effective action on such a background simply vanishes.
This reflects the supersymmetry of the rotating cylinder solution considered here.

\subsection{Rotating torus}
\label{sec:two-rot-cyl}

In order to have a 3+1-dimensional non-compact geometry,
we now consider a solution with geometry $\R^4 \times T^2 \subset \R^{10}$, embedded as
\begin{align}
X^A = \begin{pmatrix}
X^\mu\\
R_1 U_1 \\
R_2 U_2
\end{pmatrix} , 
\label{rotating-torus}
\end{align}
and algebra
\begin{align}
\co{X^\mu}{X^\nu}&=i\th^{\mu\nu}\,, &
\co{X^\mu}{U_i}&=-\xi_i^\mu U_i, & \co{U_1}{U_2} &= 0
\,.
\end{align}
This can be realized  using two commuting  fuzzy cylinders $(U_1,X^1)$ and $(U_2,X^2)$, twisted 
with a two-dimensional {\nc} plane wave $[X^\mu,X^\nu] = i \theta^{\mu\nu}, \,\mu = 0,3$
(which commutes with the cylinders). Alternatively, one can also start 
with a fuzzy torus  $U_1 U_2 = q U_2 U_1$, suitably twisted  with  4-dimensional {\nc} plane waves.
This provides  solutions of the matrix model for  suitable $\xi^\mu_i$, provided
the embedding metric of $\R^4$ has Minkowski signature.
Generalizing the previous case, a convenient basis of wave functions is given by
\begin{align}
|n_1,n_2,p\rangle &= e^{i \tilde p_\mu (X^\mu - \xi_1^\mu/2- \xi_2^\mu/2)}  U_1^{-n_1}U_2^{-n_2} \,, \nn\\
\co{U_1}{.} |n_1,n_2,p\rangle &= 2\sin(\frac{\tilde p_\mu\xi^\mu}2) |n_1-1,n_2,p\rangle \,
,\label{basis-np-4D}
\end{align}
etc.
The Laplacian acting on this basis is obtained as
\begin{align}
\Box |n_1,n_2,p\rangle &= \Big(\co{R_1 {U_1}^\dagger}{\co{R_1 {U_1}}{ . }} + \co{R_2 {U_2}^\dagger}{\co{R_2 {U_2}}{ . }}
 +\co{X^\m}{\co{X_\m}{ . }}\Big)|n,p\rangle \nn\\
&=\Big(\a_1^2+\a_2^2 + (p-n_1\xi_1 - n_2 \xi_2)\cdot (p-n_1\xi_1 - n_2 \xi_2) \Big)|n,p\rangle \nn\\
 &=   \l_{n_1,n_2}|n_1,n_2,p\rangle \nn\\
\a_i &= 2R_i \sin(\frac{\tilde p_\mu\xi_i^\mu}2)\,.
 \label{eq:energy-2-rot-cyl}  
 \end{align}
 The on-shell condition $\Box {U_i}=0$ is solved if $\xi_i\cdot \xi_i = 0$, while $\Box X^\mu = 0$ holds identically.
 Assuming this on-shell condition, the eigenvalues are
 \begin{align}
 \l_{n_1,n_2}&=\a_1^2+\a_2^2+p\cdot p - 2 n_1 \xi _1\cdot p - 2 n_2 \xi_2\cdot p
  + 2 n_1 n_2 \xi_1\cdot\xi_2
\,. \label{eq:thematrix_newbox2}
\end{align}
Now we are ready to approach the one-loop computation.

\paragraph{Bosonic modes.}

Generalizing the previous case, we note that
{\allowdisplaybreaks
\begin{align}
&\Box  + \Sigma^{(Y)}_{ab}[\Th^{ab},.]= \cU^{-1} M_Y \cU \,, \qquad
\cU= \begin{pmatrix}
 1 & 0 & 0 & 0 & 0 & 0 & 0 \\
 0 & 1 & 0 & 0 & 0 & 0 & 0 \\
 0 & 0 & 1 & 0 & 0 & 0 & 0 \\
 0 & 0 & 0 & \frac{1}{\sqrt{2}} & \frac{i}{\sqrt{2}} & 0 & 0 \\
 0 & 0 & 0 & -\frac{i}{\sqrt{2}} & -\frac{1}{\sqrt{2}} & 0 & 0 \\
 0 & 0 & 0 & 0 & 0 & \frac{1}{\sqrt{2}} & \frac{i}{\sqrt{2}} \\
 0 & 0 & 0 & 0 & 0 & -\frac{i}{\sqrt{2}} & -\frac{1}{\sqrt{2}}
\end{pmatrix} \,, \nn\\
&M_Y = \sqrt{2} R \begin{pmatrix}
\frac{\Box}{\sqrt{2} R} & 0 & 0 & - \xi^0_1 \hat{U}_1^\dagger  & i
    \hat{U}_1 \xi^0_1 & - \xi^0_2 \hat{U}_2^\dagger  & i \hat{U}_2 \xi^0_2 \\
 0 & \frac{\Box}{\sqrt{2} R} & 0 & -  \xi^1_1 \hat{U}_1^\dagger  & i
    \hat{U}_1 \xi^1_1 & - \xi^1_2 \hat{U}_2^\dagger  & i \hat{U}_2 \xi^1_2 \\
 0 & 0 & \frac{\Box}{\sqrt{2} R} & -  \xi^2_1 \hat{U}_1^\dagger  & i
    \hat{U}_1 \xi^2_1 & - \xi^2_2 \hat{U}_2^\dagger  & i \hat{U}_2 \xi^2_2 \\
   \hat{U}_1 \xi^0_1 & -  \hat{U}_1 \xi^1_1 & -  \hat{U}_1 \xi^2_1 & \frac{\Box}{\sqrt{2} R} & 0 & 0 & 0 \\
 i\xi^0_1 \hat{U}_1^\dagger  & -i \hat{U}_1^\dagger \xi^1_1 & -i \hat{U}_1^\dagger \xi^2_1 & 0 & \frac{\Box}{\sqrt{2} R} & 0 & 0 \\
   \xi^0_2 \hat{U}_2  & - \xi^1_2 \hat{U}_2  & -\xi^2_2 \hat{U}_2  & 0 & 0 & 
   \frac{\Box}{\sqrt{2} R} & 0 \\
 i \hat{U}_2^\dagger \xi^0_2 & -i \hat{U}_2^\dagger \xi^1_2 & -i \hat{U}_2^\dagger \xi^2_2 & 0 & 0 & 0 & \frac{\Box}{\sqrt{2} R}
\end{pmatrix} ,
\end{align}
where $\hat{U}_i:=[U_i,.]$.
To simplify the analysis, we 
choose $\xi^1_2=\xi^2_1=0$, $\xi^0_1=\xi^1_1$ and $\xi^0_2=\xi^2_2$ which allows to satisfy the on-shell condition. 
We then find 
}
\begin{align}
\begin{pmatrix}
\frac{\l_{n_1,n_2} '}{\sqrt{2}}  & 0 & 0 & - \a _1
   \xi^1_1 & i  \a _1 \xi^1_1 & - \a
   _2 \xi^2_2 & i  \a _2 \xi^2_2 \\
 0 & \frac{\l_{n_1,n_2} '}{\sqrt{2}}  & 0 & - \a _1
   \xi ^1_{1} & i  \a _1 \xi ^1_{1} & 0 & 0 \\
 0 & 0 & \frac{\l_{n_1,n_2} '}{\sqrt{2}}  & 0 & 0 & -
   \a _2 \xi ^2_{2} & i  \a _2 \xi ^2_{2} \\
  \a _1 \xi^1_1 & - \a _1 \xi ^1_{1} &
   0 & \frac{\l_{n_1-1,n_2} '}{\sqrt{2}}  & 0 & 0 & 0 \\
 i  \a _1 \xi^1_1 & -i  \a _1 \xi
   ^1_{1} & 0 & 0 & \frac{\l_{n_1+1,n_2} '}{\sqrt{2}}  & 0 & 0 \\
  \a _2 \xi^2_2 & 0 & - \a _2 \xi
   ^2_{2} & 0 & 0 & \frac{\l_{n_1,n_2-1} '}{\sqrt{2}}  & 0 \\
 i  \a _2 \xi^2_2 & 0 & -i  \a _2 \xi
   ^2_{2} & 0 & 0 & 0 & \frac{\l_{n_1,n_2+1} '}{\sqrt{2}} 
\end{pmatrix}
\begin{pmatrix}
a_{n_1,n_2}\\a_{n_1,n_2}\\a_{n_1,n_2}\\b^1_{n_1-1,n_2}\\c^1_{n_1+1,n_2}\\b^2_{n_1,n_2-1}\\c^2_{n_1,n_2+1}
\end{pmatrix}
=0 \,, \label{eq:rotatingtorus-bosonic-eigeneq}
\end{align}
where
\begin{align}
\l'_{n_1,n_2}=\l_{n_1,n_2}-\l \,.
\end{align}
It follows that the product of eigenvalues of $M_Y$ is given by
\begin{align}
\prod_i \l_i&=\l_{n_1,n_2}^3 \left(\l_{n_1-1,n_2} \l_{n_1,n_2-1} \l_{n_1,n_2+1}
   \l_{n_1+1,n_2}-\left(4 \a_1 \a_2 \xi^1_{1} \xi^2_{2}\right)^2\right)
\,. 
\end{align}

\paragraph{Fermionic representation.}
The computations of the fermionic part can be greatly simplified by choosing a convenient representation for the Gamma matrices.
In the rotating cylinder case the chiral representation was such a convenient choice.
Now we try to find a suitable generalization to the present higher dimensional situation.
As a starting point we use the Pauli matrices
\begin{align}
 \s_0&:=\begin{pmatrix}
        1&0\\0&1
       \end{pmatrix}
       \,, &
 \s_1&=\begin{pmatrix}
        0&1\\1&0
       \end{pmatrix}
\,,\nn\\
\s_2&=\begin{pmatrix}
       0&-i\\i&0
      \end{pmatrix}
 \,, &
 \s_3&=\begin{pmatrix}
        1&0\\0&-1
       \end{pmatrix}
       \,,
\end{align}
and the 4-dimensional Gamma matrices in the chiral representation, but renamed in the following way:
{\allowdisplaybreaks
\begin{align}
\g_0 &= \begin{pmatrix}0& 0& 1& 0\\0& 0& 0& 1\\1& 0& 0& 0\\0& 1& 0& 0\end{pmatrix}
\,, &
\g_2 &= \begin{pmatrix}0& 0& 0& 1\\0& 0& 1& 0\\0& -1& 0& 0\\-1& 0& 0& 0\end{pmatrix}
\,,\nn\\
\g_3 &=\begin{pmatrix} 0& 0& 0& -i\\0& 0& i& 0\\0& i& 0& 0\\-i& 0& 0& 0\end{pmatrix}
\,, &
\g_1 &=\begin{pmatrix} 0& 0& 1& 0\\0& 0& 0& -1\\-1& 0& 0& 0\\0& 1& 0& 0\end{pmatrix}
\,,\nn\\*
\g_5&=-i\g_0\g_1\g_2\g_3
\,.
\end{align}
Then a convenient representation for the 10-dimensional Clifford algebra can be constructed as\footnote{In fact we will only need $\G_0,\ldots,\G_6$ for the computations ahead.}
}
\begin{align}
 \G_0&=\g_0\otimes\id_8\,, &\G_1&=\g_1\otimes\id_8 \,, &\G_2&=\g_5\otimes(\s_3\otimes\g_1) \,,
 \nn\\
 \G_3&=\g_2\otimes\id_8\,, &\G_4&=\g_3\otimes\id_8 \,, &\G_5&=\g_5\otimes(\s_3\otimes\g_2) \,,
 \nn\\
 \G_6&=\g_5\otimes(\s_3\otimes\g_3) \,, &\G_7&=\g_5\otimes(\s_3\otimes i\g_0)  \,, &\G_8&=\g_5\otimes(\s_1\otimes i\id_4)\,,
 \nn\\
 \G_9&=\g_5\otimes(\s_2	\otimes i\id_4)
 \,. 
\end{align}
We then find an operator $\square\id_{32}+\Sigma^{(\psi)}_{ab}[\Th^{ab},.]$ where $U_i$, $U_j^\dagger$ entries decouple once more enabling us to compute its eigenvalues.
The explicit entries of this $32\times32$ matrix are easily computed, some of them reading:
\begin{align}
 &\square\id_{32}+\Sigma^{(\psi)}_{ab}[\Th^{ab},.]=\nn\\
 &R\left(\begin{array}{ccccccccccc}
   \frac{\square}R & \xi_2^2[U_2,.] &0&0&0&0&0&0& 2\xi_1^1[U_1,.] &0 &\ldots\\
   \xi_2^2[U_2^\dagger,.] &\frac{\square}R & 0&0&0&0&0&0&0 & 2\xi_1^1[U_1,.] &\ldots \\
   0&0 &\frac{\square}R & \xi_2^2[U_2,.] &0&0&0&0&0&0 &\ldots \\
   0&0&\xi_2^2[U_2^\dagger,.] &\frac{\square}R & 0&0&0&0&0&0 &\ldots \\
   \vdots &\vdots&\vdots&\vdots&\vdots&\vdots&\vdots&\vdots&\vdots&\vdots& \ddots
  \end{array}\right)
  \,. 
\end{align}
Using a similar ansatz as before, it is then possible to compute the product of the fermionic eigenvalues, which do \emph{not} match with the 
bosonic (and ghost) contributions.
In particular, this means that this 
solution is not supersymmetric. Although it is possible to proceed with the computation of the 
one-loop determinant, we refrain from pursuing this rather messy computation because it is not 
very illuminating.

\subsubsection*{Non-vanishing flux.}

Finally, it is straightforward to generalize the above solution $\R^4 \times T^2 $ \eq{rotating-torus} such that the torus has a flux, i.e.
\begin{align}
 U_1 U_2 &=q U_2 U_1, \qquad q = e^{i \theta} \ 
 \,.
\end{align}
Then the eigenvalues of the matrix Laplacian
\eq{eq:energy-2-rot-cyl} acquire additional terms,
\begin{align}
\Box |n_1,n_2,p\rangle 
&=\Big(\a_1^2+\a_2^2 + (p-n_1\xi_1 - n_2 \xi_2)\cdot (p-n_1\xi_1 - n_2 \xi_2) \Big)|n,p\rangle \nn\\
 &=   \l_{n_1,n_2}|n_1,n_2,p\rangle \,, \nn\\
\a_1 &= 2R_1 \sin\Big(\frac 12(n_2\theta-\tilde p_\mu\xi_1^\mu)\Big)\,, \qquad \a_2 = 2R_2 \sin\Big(\frac 12(n_1\theta +\tilde p_\mu\xi_2^\mu)\Big)
 \,, \label{eq:energy-2-rot-cyl-flux}  
 \end{align}
i.e. $\a_{1,2}$ have changed compared to \eq{eq:energy-2-rot-cyl}.
The on-shell condition $\Box U_i = 0$ now implies 
 \begin{align}
  \xi_i\cdot \xi_i = -4 \left(\e_{ij} R_j\right)^2  \sin^2\!\left(\theta/2\right)
  \,, \label{eq:onshell-cond}
 \end{align}
which can be solved due to the Minkowski signature on $\R^4$. 
This typically leads to massive Kaluza-Klein modes, for sufficiently large flux $\theta$ on $T^2$.
To see this, we solve the on-shell condition $\Box |n_1,n_2,p\rangle = 0$ for the KK modes, which 
for $\eta_{\mu\nu}= \diag(-1,1,1,1)$ gives
\begin{align}
(p^0-n_1\xi_1^0 - n_2 \xi_2^0)^2 &= \d_{ij} (p-n_1\xi_1 - n_2 \xi_2)^i (p-n_1\xi_1 - n_2 \xi_2)^j 
 + \a_1^2+\a_2^2 
 \,. \label{on-shell-torus-explicit}
\end{align}
Consider the simplest solution with $\xi_i^\mu = (\xi_i^0,0,0,0)$, so that
$(\xi_i^0)^2 \approx (\e_{ij}R_j)^2\theta^2$. Then the 
time-like component $p^0$ does not get any contribution from the 
arguments  $\tilde p_\mu\xi^\mu_i = p^\mu \theta^{-1}_{\mu\nu}\xi^\nu$ in 
$\a_{i}$ (assuming that $\theta^{\mu\nu}$ is non-degenerate on $\R^4$), 
and the above on-shell relation becomes
\begin{align}
(p^0-n_1\xi_1^0 - n_2 \xi_2^0)^2 
 \ &\sim  \  g_{ij} (p^i -  A^i) (p^j - A^j) + m^2  
\end{align}
for some effective 3-metric  $g_{ij}$ and 3-vector $A^i$.
Here the effective mass $m^2$ is manifestly positive since all terms on the rhs of \eq{on-shell-torus-explicit}
are positive (note that the mass shell might be deformed).
\begin{figure}[ht]
\centering
\def\svgwidth{\columnwidth}
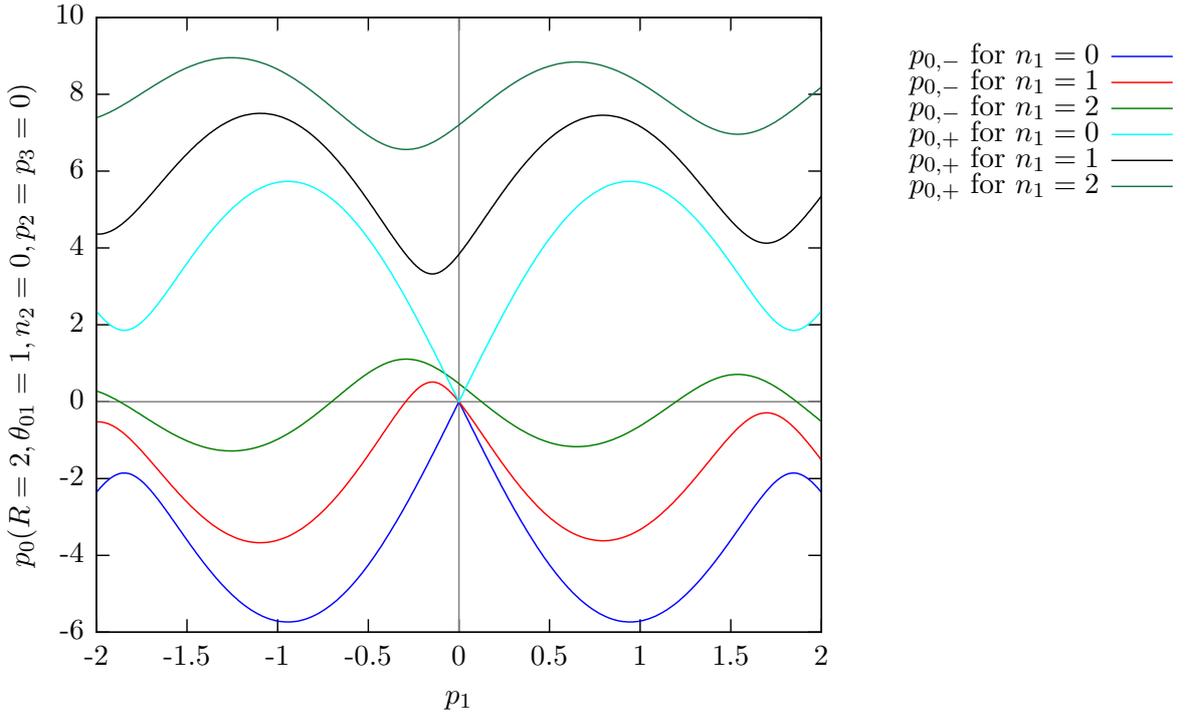
\caption{Energy levels, with flux}
\label{fig:energy-with-flux}
\end{figure}
Therefore there  is an energy gap $m^2 > 0$, 
albeit the effective mass shells for the non-trivial KK modes are typically shifted in 4-momentum space.
Similar band structures are familiar from  solid state physics. 
In particular, this suggests that the system can be quantized consistently, and 
the one-loop corrections should then be UV finite as discussed above.
On the other hand, due to the shifts in momentum space 
at least one positive energy branch intersects the negative energy branch of 
another KK mode, see \figref{fig:energy-with-flux}.
Specifically, this happens 
for the $n_i=(0,1)$ and the $n_i=(1,0)$ mode\footnote{The crossing of the corresponding energy shells at $p^\mu = 0$ 
will lead to mixing of the corresponding modes. This corresponds to the mixing 
of gravitational modes found in \cite{Steinacker:2012ra}.}, 
which are  on-shell for $p^\mu = 0$ since $\Box U_i = 0$ by construction. 
However,  these modes have distinct quantum numbers $n_i$ which are conserved
by the interactions, so that such an intersection of mass shells does not 
necessarily imply an instability. Moreover, the above solution might get deformed 
into more stable solutions upon taking the interactions into account.

In any case, it appears likely that a stable,
non-trivial solution can be found  due to the  non-trivial flux on $T^2$, 
however more  work is required to settle this issue.
It should also be pointed out that in the presence of flux,
the torus can be realized in terms of a fuzzy torus  $T^2_N$, leading to 
a truncation of the KK modes; this case is also studied in \cite{Polychronakos:2013fma}.

For the solutions without flux, the stability analysis is more complicated 
because the arguments of $\a_{i}$ necessarily contain $p^0$. 
Although preliminary investigations suggest that this case also 
leads to a positive mass gap, a more detailed investigation is required which is left 
for future work.

\section{Conclusion and Outlook}
In this work we have studied some compactified brane solutions of type $\R^2 \times S^1$ and $\R^4 \times T^2$ 
in the IKKT  model, the latter with as well as without flux on $T^2$.
We determined the full spectrum of bosonic and fermionic modes on such a background, as required 
for the 1-loop quantization.
The one-loop effective action is manifestly UV finite on branes $\dim\cM \leq 8$, due to the 
maximal supersymmetry of the IKKT model.
Although a general criterion is known \cite{Ishibashi:1996xs}, in practice it is not easy to decide for 
 which types of compactifications preserve or break 
supersymmetry.
We found that the geometry $\R^2 \times S^1$ leads to a supersymmetric spectrum, while 
the $\R^4 \times T^2$ does not, 
leading to a complicated spectrum of bosonic and fermionic Kaluza-Klein (KK)  modes.
The higher KK modes are typically shifted in 4-momentum space
 due to the rotation of $T^2$, leading to a complicated pattern of KK modes 
which prevents an explicit evaluation of  the one-loop effective action.
We also  verified that the KK modes have a positive or vanishing mass gap.
However, certain different KK modes have an intersecting mass shell due to a shift in 
momentum space, which may or may not indicate an instability.
To settle the issue of stability in the presence of interactions requires further work.

\subsection*{Acknowledgements}
D.N. Blaschke is a recipient of an APART fellowship of the Austrian Academy of Sciences, and is also grateful for the hospitality of the theory division of LANL and its partial financial support.
The work of H.S. is supported by the Austrian Fonds f\"ur Wissenschaft und Forschung under grant P24713.


\bibliographystyle{../custom1.bst}
\bibliography{../articles.bib,../books.bib}

\end{document}

%% file: withflux.pdf_tex
\begingroup%
  \makeatletter%
  \providecommand\color[2][]{%
    \errmessage{(Inkscape) Color is used for the text in Inkscape, but the package 'color.sty' is not loaded}%
    \renewcommand\color[2][]{}%
  }%
  \providecommand\transparent[1]{%
    \errmessage{(Inkscape) Transparency is used (non-zero) for the text in Inkscape, but the package 'transparent.sty' is not loaded}%
    \renewcommand\transparent[1]{}%
  }%
  \providecommand\rotatebox[2]{#2}%
  \ifx\svgwidth\undefined%
    \setlength{\unitlength}{656bp}%
    \ifx\svgscale\undefined%
      \relax%
    \else%
      \setlength{\unitlength}{\unitlength * \real{\svgscale}}%
    \fi%
  \else%
    \setlength{\unitlength}{\svgwidth}%
  \fi%
  \global\let\svgwidth\undefined%
  \global\let\svgscale\undefined%
  \makeatother%
  \begin{picture}(1,0.6097561)%
    \put(0,0){\includegraphics[width=\unitlength]{withflux.pdf}}%
    \put(0.06743902,0.0647561){\makebox(0,0)[rb]{\smash{-6}}}%
    \put(0.06743902,0.12963415){\makebox(0,0)[rb]{\smash{-4}}}%
    \put(0.06743902,0.1945122){\makebox(0,0)[rb]{\smash{-2}}}%
    \put(0.06743902,0.25939024){\makebox(0,0)[rb]{\smash{0}}}%
    \put(0.06743902,0.32439024){\makebox(0,0)[rb]{\smash{2}}}%
    \put(0.06743902,0.38926829){\makebox(0,0)[rb]{\smash{4}}}%
    \put(0.06743902,0.45414634){\makebox(0,0)[rb]{\smash{6}}}%
    \put(0.06743902,0.51902439){\makebox(0,0)[rb]{\smash{8}}}%
    \put(0.06743902,0.58390244){\makebox(0,0)[rb]{\smash{10}}}%
    \put(0.07756098,0.04280488){\makebox(0,0)[b]{\smash{-2}}}%
    \put(0.15353659,0.04280488){\makebox(0,0)[b]{\smash{-1.5}}}%
    \put(0.2295122,0.04280488){\makebox(0,0)[b]{\smash{-1}}}%
    \put(0.3054878,0.04280488){\makebox(0,0)[b]{\smash{-0.5}}}%
    \put(0.38146341,0.04280488){\makebox(0,0)[b]{\smash{0}}}%
    \put(0.45731707,0.04280488){\makebox(0,0)[b]{\smash{0.5}}}%
    \put(0.53329268,0.04280488){\makebox(0,0)[b]{\smash{1}}}%
    \put(0.60926829,0.04280488){\makebox(0,0)[b]{\smash{1.5}}}%
    \put(0.6852439,0.04280488){\makebox(0,0)[b]{\smash{2}}}%
    \put(0.02146341,0.3297561){\rotatebox{90}{\makebox(0,0)[b]{\smash{$p_0(R=2,\theta_{01}=1,n_2=0,p_2=p_3=0)$}}}}%
    \put(0.38134146,0.00987805){\makebox(0,0)[b]{\smash{$p_1$}}}%
    \put(0.91804878,0.55097561){\makebox(0,0)[rb]{\smash{$p_{0,-}$ for $n_1=0$}}}%
    \put(0.91804878,0.52902439){\makebox(0,0)[rb]{\smash{$p_{0,-}$ for $n_1=1$}}}%
    \put(0.91804878,0.50707317){\makebox(0,0)[rb]{\smash{$p_{0,-}$ for $n_1=2$}}}%
    \put(0.91804878,0.48512195){\makebox(0,0)[rb]{\smash{$p_{0,+}$ for $n_1=0$}}}%
    \put(0.91804878,0.46317073){\makebox(0,0)[rb]{\smash{$p_{0,+}$ for $n_1=1$}}}%
    \put(0.91804878,0.44121951){\makebox(0,0)[rb]{\smash{$p_{0,+}$ for $n_1=2$}}}%
  \end{picture}%
\endgroup%